\documentclass[twocolumn,english,aps,prl,reprint,floatfix,notitlepage,footinbib,superscriptaddress]{revtex4-1}
\pdfoutput=1

\usepackage{amsmath}
\usepackage{physics}
\usepackage{lipsum}

\usepackage{bbold}

\usepackage[normalem]{ulem}
\usepackage[T1]{fontenc}
\usepackage[latin9]{inputenc}
\usepackage{geometry}
\usepackage{amsfonts}
\usepackage{geometry}
\usepackage{comment}
\usepackage{mathtools}
\usepackage{float}
\usepackage{slashed}
\usepackage{ragged2e}
\usepackage{array}
\usepackage{bbold}
\usepackage{tikz}

\makeatletter\g@addto@macro\bfseries{\boldmath}\makeatother

\usepackage{stackengine}
\usepackage{esint}
\usepackage{breakurl}
\usepackage{breakurl}
\usepackage[hang,flushmargin]{footmisc} 
\allowdisplaybreaks
\makeatletter

\usepackage{etoolbox}
\apptocmd{\thebibliography}{\justifying\setlength{\leftskip}{7.4mm}}{}{} 
 
 \usepackage{relsize}
\usepackage{babel}

\makeatletter
\def\simgt{\mathrel{\lower2.5pt\vbox{\lineskip=0pt\baselineskip=0pt
           \hbox{$>$}\hbox{$\sim$}}}}
\def\simlt{\mathrel{\lower2.5pt\vbox{\lineskip=0pt\baselineskip=0pt
           \hbox{$<$}\hbox{$\sim$}}}}
\makeatother

\usepackage{changepage}

\newcommand{\Od}[1]{\mathcal{O}{\left(#1\right)}}

\begin{document}

\title{Entropy from scattering in weakly interacting systems}

\author{Duncan MacIntyre$^{a,b}$ and Gordon W. Semenoff$^a$\\~~\\
$^a$ Department of Physics and Astronomy, University of British Columbia,\\
 ~~6224 Agricultural Road, Vancouver, British Columbia, Canada V6T 1Z1\\
 $^b$ 
 Department of Physics, University of Toronto, 60 St.~George St.,~~~~~~~~~~ \\
 Toronto, Ontario, M5S 1A7, Canada
}
 
\begin{abstract}
\noindent 

Perturbation theory is used to investigate the evolution of the von Neumann entropy of a subsystem of a bipartite quantum system under the action of a unitary matrix, in the limit where that matrix is close to the unit matrix.  The physical context for such process would be scattering with weak short-ranged interactions where the unitary matrix is the S matrix.  We find surprisingly simple criteria for the initial state and the S matrix  that guarantee that the subsystem entropy  increases. The class of initial states that meet these criteria are more correlated than simple product states of the subsystems.  They form a subclass  of the set of all separable states, and they can therefore be assembled by classical processes alone.
  
\end{abstract}
\maketitle

\preprint{}
\date{}

\maketitle

\section{Introduction}

Amongst the ``area laws'' of fundamental physics,  a rather interesting one  has emerged in the context of scattering theory.  The question which is asked concerns the quantum entanglement between subsets of the particle degrees of freedom that is produced by interactions in the course of a scattering event.  It considers a bipartite quantum system that is composed of two subsystems, A and B, and with the  incoming particles in an unentangled pure quantum state.  The interaction of the particles is assumed to be weak, so that perturbation theory is valid. Then the change in von Neumann entropy of one of the subsystems per unit time and per unit of incoming beam flux  is  proportional to the total cross-section, \(\sigma\), which is measured in units of area \cite{seki1,Grignani:2016igg,Fan:2017mth,Boyanovsky:2018fxl,Peschanski:2019yah,Low:2024mrk,Low:2024hvn}.
We write
\begin{align}
\label{area law}
\frac{\delta S^A}{{\rm (time)(beam ~flux)}}~=~\left[\lambda^2\ln\frac{1}{\lambda^2}\right] \sigma
\end{align}
The parameter $\lambda$ is the coupling constant which is assumed to be of sufficiently small magnitude that perturbation theory is valid.   The appearance of its logarithm in the above expression is interesting.  It is due to the singular nature of von Neumann entropy in this limit. 

 In equation (\ref{area law}), the entropy of the A subsystem is increased by the scattering process since the cross section is necessarily positive.  This is not surprising.  Because the incoming unentangled pure state has zero entanglement entropy and because entropy is non-negative, the entropy can only increase.  
 
 In this paper we will study situations where the incoming state is more general than an unentangled pure state.  Generally, it is specified by an incoming density matrix $\hat{\rho}^{\rm in}$.  We will find criteria for this incoming state and for the scattering matrix so that the entropy of the A subsystem is guaranteed to be increasing. In studying this issue, we will derive a refinement of the formula (\ref{area law}) for more general incoming states. We will also consider some examples which demonstrate that, when our positivity criteria are not met, that subsystem entropy can either increase or decrease, depending on the fine details of the initial state.  Although we will not explore the subject here, our results should be pertinent to the discussions of a ``second law of entanglement'' \cite{T}.  
 
  Our work will study evolution of a bipartite quantum system when its state is acted on with a unitary matrix, in the limit where that matrix is close to the unit matrix.    While, as we have done, it is very natural to cast the problem which we address in the context of scattering, where degrees of freedom in an initial state come together and interact over a limited time interval before flying apart and becoming final states and where, if the interactions are weak, the S matrix is indeed close to the identity,  the results are more general and they apply to any bipartite quantum system to which one would apply such a ``small'' unitary gate.  In that case the question would simply ask what class of states of a bipartite system have non-increasing subsystem entropy when they are evolved by such a gate.

 A recent interesting work \cite{Cheung:2023hkq}  studied the case of an incoming state which is still an unentangled product state of the subsystems but where each subsystem can be in a mixed state, so that the incoming state is a  direct product of density matrices of the subsystems, $\hat{\rho}=\hat{\rho}^A\otimes\hat{\rho}^B$.   They asked which initial states of this kind must always have non-decreasing  $n$-Tsallis entropy (for $n\geq 2$) of 
subsystem A, no matter what the interaction as long as time evolution is unitary, and no matter what the initial state $\hat{\rho}^B$ of subsystem B. 
They found a surprisingly simple characterization of such states:
the $n$-Tsallis entropy of subsystem $A$ is always non-decreasing at the second order in perturbation theory if $\hat{\rho}^A$ is proportional to a projection operator, that is, if $\hat{\rho}^A$ has finitely many non-zero eigenvalues and  all of the non-zero eigenvalues are equal.   

 In the following we will ask a similar question for the von Neumann entropy.  We will assume a more general initial state which is not necessarily a product of the states of the two subsystems.  We will retain the assumption of  unitarity, in that the $\hat{T}$ matrix satisfies the optical theorem and, as in reference  \cite{Cheung:2023hkq},  we will ignore any possible constraints due to causality, locality, or spacetime and internal symmetries.  In a sense, the arbitrary $\hat{T}$ matrix will be our Maxwell's demon and our goal is to understand how to organize the initial states of the system in such a way that we are guaranteed to thwart the demon's efforts to decrease the entropy of subsystem A.  
 
We will find that the von Neumann entropy of subsystem A is non-decreasing for a class of initial states which are also easy to characterize.  Consider the incoming reduced density matrix of subsystem A,  $\hat{\rho}^{A{\rm in}}$, its eigenvectors $\ket{m}$,  and its eigenvalues $\rho^{A{\rm in}}_m$, with
 \begin{equation} \begin{aligned}
 &\hat{\rho}^{A{\rm in}}=\sum_m \ket{m}\bra{m}~\rho^{A{\rm in}}_{m}
 \\
 &{\rm Tr}\left[\hat{\rho}^{A{\rm in}}\right]= \sum_m\rho^{A{\rm in}}_{m}=1,~
 \rho^{A{\rm in}}_{m}\geq 0.
 \label{properties}
 \end{aligned}
 \end{equation}
We will find that, for sufficiently weak interactions, the von Neumann entropy contained in $\hat{\rho}^{A{\rm in}}$ is guaranteed to be increasing when
 \begin{enumerate}
 \item{} $\hat{\rho}^{A{\rm in}}$ has a nonempty kernel, that is,
 $$
 \exists \ket{m} ~{\rm such~that~}\hat{\rho}^{A{\rm in}}\ket{m}=0;
 $$
 \item{}
For any eigenvector $ \ket{m}$ of  $\hat{\rho}^{A{\rm in}}$,  
\begin{equation} \begin{aligned}
\biggl[ { \ket{m}}\bra{m}\otimes{\mathbb{1}}^B,\; \hat{\rho}^{\rm in}\biggr]=0; \hspace{3em}\text{and}
\label{separable0}\end{aligned}\end{equation}
\item{}
The the leading term in the perturbative expansion of the \(\hat{T}\) matrix, which we shall denote as \(\lambda\hat{T}^{(1)}\), 
allows for transitions between the non-kernel and the kernel states of $\hat{\rho}^{A{\rm in}}$ and it acts nontrivially on the B subsystem.
\end{enumerate}

When the first two of the three criteria listed above are both satisfied, the change in entropy is at least of order \(\lambda^2 \ln \frac{1}{\lambda^2}\) as \(\lambda \to 0\) and the change in entropy is guaranteed to be non-negative at order \(\lambda^2 \ln \frac{1}{\lambda^2}\). 

The change in entropy could still vanish at order \(\lambda^2 \ln \frac{1}{\lambda^2}\) and it would not necessarily be non-negative at higher orders. The third criterion above is required for the change in entropy to be nonzero at order \(\lambda^2 \ln \frac{1}{\lambda^2}\). When all three criteria are satisfied, the change in entropy is strictly positive.
 
 Equation (\ref{separable0}) is equivalent to the criterion that there exist bases of states $\ket{m}$ of the A subsystem and $\ket{\tilde m}$ of the B subsystem so that, in the space spanned by the product states $\ket{m,\tilde m}\equiv\ket{m}\otimes\ket{\tilde m}$, the incoming density matrix has the form
 \begin{equation} \label{special0}
 \hat{\rho}^{\rm in}=\sum_{m,\tilde m}\ket{m,\tilde m}\bra{m,\tilde m}~\rho^{\rm in}_{m,\tilde m}
 \end{equation}
 for some \(\rho^{\rm in}_{m,\tilde m}\in[0,1]\) with \(\sum_{m\tilde m}\rho^{\rm in}_{m\tilde m}=1\). In this case
\(
\rho^{A{\rm in}}_m = \sum_{\tilde m}\rho^{\rm in}_{m,\tilde m}
\).

Density matrices with the property (\ref{separable0}) or (\ref{special0}) are 
more correlated than direct product states.  They are a subclass of the set of separable states.
Separable states are mixed states of a bipartite quantum system which are distinguished by the fact that 
   the correlations in such a state can be built up by classical processes alone.  
  They are defined as any states whose density matrices are  convex combinations of product states \cite{toth},
   \begin{align}\label{separable}
  \hat \rho_{\rm separable} = \sum_i k_i \hat\rho_i^A\otimes\hat \rho_i^B
   ~,~k_i>0,~\sum_ik_i=1,
   \end{align}
  where $\hat\rho_i^A$ and $\hat\rho_i^B$ are Hermitian matrices with the positivity and normalization properties of density matrices. 
  
    It is clear that states which obey (\ref{separable0}) and therefore have the form (\ref{special0}) are a subclass of the set of separable states (\ref{separable}) where the matrices in the sum in (\ref{separable}) commute with each other, $[\hat\rho_i^A\otimes \hat\rho_i^B,\hat\rho_j^A\otimes \hat\rho_j^B]=0$ $\forall i,j$, and can therefore be simultaneously diagonal.

 The projection operator states of reference \cite{Cheung:2023hkq}, for which the n-Tsallis entropy is non-decreasing,  are certainly in the class of states obeying (\ref{separable0}) or (\ref{special0}), as are other states.  The reason why the set of states obeying (\ref{separable0}) or (\ref{special0}), where entropy cannot decrease, is larger for von Neumann entropy than it is for n-Tsallis entropy with $n>2$ is the singular nature of the  perturbative expression for the von Neumann entropy which has dominant terms that are different from those in the n-Tsallis entropy. This is already reflected in the logarithm of the coupling constant in the area law (\ref{area law}) and we shall find a more general expression for the change in entropy of more general incoming states, with the same $\left[\lambda^2\ln\frac{1}{\lambda^2}\right]$ coefficient.

 \section{Perturbative scattering}
 
To facilitate scattering, an incoming quantum state, which we will describe by the density matrix $\hat{\rho}^{\rm in}$,  is prepared at an early initial time. It could, for example, contain a number of free relativistic particles whose initial state is characterized by $\hat{\rho}^{\rm in}$.  Then, the particles encounter each other, scatter, and the state evolves  to an outgoing state described by a density matrix $\hat{\rho}^{\rm out}$ at a late time.  The outgoing state  is related to the incoming state  by the \(\hat{S}\) matrix in that
\begin{align}\label{s matrix}
 \hat{\rho}^{\rm out} = \hat{S}\hat{\rho}^{\rm in} \hat{S}^\dagger.
\end{align}
We will assume that the $\hat S$ matrix is unitary, $\hat{S}\hat{S}^\dagger=\mathbb{1}$. 
The $\hat{T}$ matrix is defined so that
\begin{align}
\hat{S}=\mathbb{1} +i \hat{T}
\end{align}
The unitarity of $\hat{S}$  is equivalent to 
\begin{align}\label{optical theorem}
i\hat{T}-i\hat{T}^\dagger =- \hat{T}\hat{T}^\dagger
\end{align}
which is the optical theorem.

We assume that the  
$\hat{T}$ matrix has an asymptotic expansion  in a small, dimensionless parameter $\lambda$
\begin{align}\label{te}
\hat{T} &= \sum_{k=1}^\infty \lambda^k \hat{T}^{(k)},
&
&\lambda \to 0.
\end{align} 
   
Equation (\ref{optical theorem}) and  (\ref{te}) 
give the evolution of the density matrix to order $\lambda^2$ as
\begin{align}
 \hat{\rho}^{\rm out}=\,&(\mathbb{1} +i \hat{T})\hat{\rho}^{\rm in} (\mathbb{1} -i \hat{T}^\dagger)~= \hat{\rho}^{\rm in}+\delta\hat{\rho}, \nonumber \\
\delta\hat{\rho}=\,&
i\lambda\left[  \hat{T}^{(1)},\hat{\rho}^{\rm in} \right] +i\frac{\lambda^2}{2}\left[
 \hat{T}^{(2)}+\hat{T}^{(2)^\dagger},\hat{\rho}^{\rm in} \right]
\nonumber \\
&
~~~-\frac{\lambda^2}{2}\left[ \hat{T}^{(1)} , \left[ \hat{T}^{(1)},\hat{\rho}^{\rm in} \right]\right]
+\Od{\lambda^3}.
\label{perturbative final density matrix}
\end{align}
(The optical theorem (\ref{optical theorem}) implies that $\hat{T}^{(1)}= \hat{T}^{(1)\dagger}$ and $\hat{T}^{(2)}-\hat{T}^{(2)\dagger}=i\hat{T}^{(1)}\hat{T}^{(1)}$.)

The initial and final density matrices must be non-negative and 
normalized,
${\rm Tr}[\hat{\rho}^{\rm in}]=1$, ${\rm Tr}[\hat{\rho}^{\rm out} ]=1$, and ${\rm Tr}[\delta\hat{\rho}]=0$.
 Since $\hat{\rho}^{\rm in}$ and $\hat{\rho}^{\rm out}$  are related by the unitary conjugation in equation  (\ref{s matrix}), they must have identical eigenvalues.   However, this is not so for the initial and final reduced density matrices $\hat{\rho}^{A{\rm in}}$  
 and  $\hat{\rho}^{A{\rm out}}$  and a study of the evolution of the eigenvalues 
 of $\hat{\rho}^{A{\rm in}}$ will be our main tool in the following discussion.  
 
We will be interested in the change of the von Neumann entropy of subsystem A.  This quantity 
can be expressed in terms of the eigenvalues \(\rho^{A{\rm in}}_{m}\) and \(\rho^{A{\rm out}}_{m}\) of its reduced incoming and outgoing density matrices. We compute
\begin{align}
\nonumber
\delta S^A &= S^A_{\rm out}-S^A_{\rm in}\\\nonumber
&=-{\rm Tr}\biggl[\hat{\rho}^{A{\rm out}}\ln\hat{\rho}^{A{\rm out}}\biggr]+{\rm Tr}\biggl[\hat{\rho}^{A{\rm in}}\ln\hat{\rho}^{A{\rm in}}\biggr]
\\
&
=-\sum_m \left[ \rho^{A{\rm out}}_{m}\ln\rho^{A{\rm out}}_{m}
-  \rho^{A{\rm in}}_{m}\ln\rho^{A{\rm in}}_{m} \right].
\label{vn entropy2}
\end{align}
To study the evolution of the eigenvalues,  
we must begin with the evolution of the 
full density matrix in equation (\ref{perturbative final density matrix}). By taking its partial trace over the $\ket{\tilde m}$ states of the B subsystem,  we obtain the expansion
 \begin{equation}
 \delta \hat{\rho}^A = \hat{\rho}^{A{\rm out}} - \hat{\rho}^{A{\rm in}} =\lambda \delta \hat{\rho}^{A(1)} +\lambda^2 \delta \hat{\rho}^{A(2)}+\Od{\lambda^3} \label{delta lambda}.
\end{equation}
Finding the shifts of the  eigenvalues  
 is an exercise in perturbation theory where, for non-degenerate eigenvalues, we must find the appropriate diagonal elements of the matrix 
\begin{align}
\mathcal M_{mm'}=\,&\lambda\bra{m}\delta \hat{\rho}^{A(1)}\ket{m'}+\lambda^2\bra{m}\delta \hat{\rho}^{A(2)}\ket{m'}
\nonumber
\\ &+\lambda^2 \sum_{m''\neq m}\frac{\bra{m} \delta\rho^{A(1)}\ket{m''}\bra{m''}\delta\rho^{A(1)} \ket{m'} }{ \rho^A_{m}-\rho^A_{m''} } \nonumber \\ &+\Od{\lambda^3}
\label{matrix}
\end{align}
and for degenerate eigenvalues we must find the eigenvalues of $\mathcal M_{mm'}$ with $m,m'$ restricted
to the degenerate eigenspace.  In that case, $m''$ in the last term in (\ref{matrix}) must be summed over
those states which are not   degenerate with the ones labeled by $m$ and $m'$.

Putting in equation (\ref{perturbative final density matrix}), we find
\begin{align}\label{shift of the kernel}
&\mathcal M_{mm'}=
i\lambda \sum_{\tilde m} \bra{m,\tilde m}\left[  \hat{T}^{(1)},\hat{\rho}^{\rm in}\right] \ket{m',\tilde m}
 \\\nonumber & 
+\lambda^2\sum_{\tilde m}\biggl( \frac{i}{2}\bra{m,\tilde m}\left[
 \hat{T}^{(2)}+\hat{T}^{(2)^\dagger},\hat{\rho}^{\rm in}\right]\ket{m',\tilde m}
 \\ \nonumber& 
- \frac{1}{2} \bra{m,\tilde m} \left[\hat{T}^{(1)},\left[  \hat{T}^{(1)}, \hat{\rho}^{\rm in}\right]\right]\ket{m',\tilde m} ~-
 \\
 \nonumber &
  \sum\limits_{ \substack{{m''}\\ \tilde m'} }  
\frac{\bra{m,\tilde m} [\hat{T}^{(1)},\hat{\rho}^{\rm in}] \ket{m'',\tilde m}
\bra{m'',\tilde m'}[\hat{T}^{(1)},\hat{\rho}^{\rm in}] \ket{m',\tilde m'}
  }{ \rho^{A{\rm in}}_{m}- \rho^{A{\rm in}}_{m''} }
  \biggr)
 \\
 \nonumber &
+\Od{\lambda^3}.
\nonumber
\end{align}


\section{The leading order $\sim\lambda$}

Let us first examine the contribution of the leading, linear term in $\lambda$ in equation (\ref{shift of the kernel}).  If this linear term  is nonzero it will dominate the shift of the eigenvalues in the limit $\lambda\to 0$. 

It will be useful to separate the states \(\ket{m}\) of the A subsystem into those states which have nonzero and those which have zero probability of appearing in the incoming state.  The latter are in the kernel of $\hat{\rho}^{A{\rm in}}$ and have indices in
\begin{equation}\begin{aligned}\label{kernel}
\kappa = \left\{m~\biggl|~\hat{\rho}^{A{\rm in}}\ket{m}=0\right\}.
\end{aligned}\end{equation}

Let us begin with states in the kernel of $\hat{\rho}^{A{\rm in}}$. Since the reduced density matrix must always be non-negative, the shift of each of
its zero eigenvalues must be non-negative. Thus $\delta\rho^A_{m}\geq 0$ whenever \(m \in \kappa\).  These shifts of eigenvalues are the eigenvalues of the matrix $\mathcal M$ given in equation (\ref{shift of the kernel}) restricted to \(m, m' \in \kappa\).

But if any of these eigenvalues are nonzero at all, their sign can be flipped by simply flipping the sign of $\hat{T}^{(1)}$ which flips the sign of the linear term in $\mathcal M$.  Thus, if  $\delta\rho^A_{m}\neq 0$ at linear order in $\lambda$ for some \(m \in \kappa\), we can always engineer a $\hat{T}^{(1)}$ so that  $\delta\rho^A_{m}<0$ which is a contradiction.  The only way to avoid this contradiction is for all of these shifts of eigenvalues to be zero.  
It must be that $\mathcal M_{mm'}=0$ at the linear order in \(\lambda\) for all $m,m'\in\kappa$, and for any $\hat{T}^{(1)}$.
This is equivalent to the condition
\begin{equation}
\begin{aligned}
& \biggl[~ \ket{m}\bra{m'}\otimes\mathbb{1}^B~,~\hat{\rho}^{\rm in}~\biggr] =0 ~~\forall m,m'\in \kappa
\label{rho ker}
\end{aligned}
\end{equation}
 where $\mathbb{1}^B$ is the unit matrix acting on the  $\ket{\tilde m}$ states.
 Note that, so far,  we have only used the fact that $\hat{\rho}^{\rm in}$ and $\hat{\rho}^{\rm in}+\delta\rho$ are density matrices.  For any acceptable density matrix, equation (\ref{rho ker}) must hold.

Now that we understand that the zero eigenvalues of $ \hat{\rho}^{A{\rm in}}$ 
cannot change at all to linear order in $\lambda$, we are left with the shifts of the
non-zero eigenvalues which, for non-degenerate $\rho^{A{\rm in}}_{m}$, are given by the diagonal elements $\mathcal M_{mm}$ in (\ref{shift of  the kernel}) for $m\notin\kappa$ and for degenerate $\rho^{A{\rm in}}_{m}$ are gotten by diagonalizing $\mathcal M_{mm'}$ in the degenerate subspace.  

Using (\ref{vn entropy2}), and using the fact that $\delta\rho_{m}^{A(1)}=0$ for \(m \in \kappa\), we find that the change in subsystem entropy is
\begin{equation}
\delta S^A =
 -\lambda \sum_{m\notin\kappa} \delta\rho_{m}^{A(1)}~\left[\ln\rho^{A{\rm in}}_{m}+1\right]+\Od{\lambda^2\ln\frac{1}{\lambda^2}}.
\label{first order delta S}
\end{equation}
We can find $\delta\rho_{m}^{A(1)}$ from the first term on the right-hand-side of equation (\ref{shift of the kernel}).  In the case of a degenerate eigenvalue, it is given by the eigenvalues of the matrix 
\begin{equation}\begin{aligned}
&\delta \rho_{mm'}^{A(1)} =  i {\rm Tr}\left(\hat{T}^{(1)}\left[ \hat{\rho}^{\rm in}, \ket{m}\bra{m'}\otimes\mathbb{1}^B\right] 
\right)
\end{aligned}\label{first order delta S1}
\end{equation}
where we restrict \(m\) and \(m'\) to the degenerate subspace with \(\rho_m^A = \rho_{m'}^A\).

Given that $\delta \rho_{mm'}^{A(1)} $ is itself  a Hermitian matrix, 
 if (\ref{first order delta S1}) has any terms which are not zero, 
 at lease some of the eigenvalues of $\delta \rho_{mm'}^{A(1)} $ would be non-zero and 
 the right-hand-side of (\ref{first order delta S}) would non-zero.
Again, since we can flip the sign of $\hat{T}^{(1)}$ at will, if the right-hand-side of (\ref{first order delta S}) is non-zero, we can always engineer a time evolution where the entropy can decrease.  
The only way to prevent this possibility is choose the initial state $\hat{\rho}^{\rm in}$ so that the linear contribution to the shift in entropy in equation (\ref{first order delta S}) vanishes. 
From equation  (\ref{first order delta S}), and together with (\ref{rho ker}),
this requires
\begin{align}\label{separable1}
&\biggl[~ \ket{m}\bra{m'}\otimes\mathbb{1}^B~,~  \hat{\rho}^{\rm in}~\biggr] =0 \\
&\text{for all }m,m'\text{ with }\rho_m^A = \rho_{m'}^A.
\nonumber
\end{align}
A close examination reveals that (\ref{separable0}), (\ref{special0}), and (\ref{separable1}) are all equivalent.

We  conclude that, if the initial state satisfies (\ref{separable0}), (\ref{special0}), or (\ref{separable1}), the linear term
in $\lambda$ in the time evolution of the entropy of subsystem A vanishes. 
In this case, the change in entropy will be of order \(\lambda^2 \ln\frac{1}{\lambda^2}\) or higher and we must use second-order perturbation theory.

  \section{Second order $\sim\lambda^2\ln(1/\lambda^2)$}
  
  We shall now assume that the initial state obeys  (\ref{separable0}) and is of the form (\ref{special0}) so that the leading contribution to the shift in eigenvalues of the reduced density matrix is obtained by second order perturbation theory. 
  
   If the initial reduced density matrix $\hat{\rho}^{A{\rm in}}$ 
has a nonempty kernel, the zero eigenvalues contribute a non-analytic term in $\lambda$ that will dominate
in the limit $\lambda \to 0$. We have
\begin{align} \label{2ndorder}
& \delta S= \left[\lambda^2\ln\frac{1}{\lambda^2}\right]  \\ &
\times \sum\limits_{ \substack{m\notin\kappa\\m'\in\kappa \\ \tilde m,\tilde m'}}   \biggl(  \rho^{{\rm in}}_{m,\tilde m}  \left|{T}^{(1)}_{\substack{mm'\\\tilde m\tilde m'}}\right|^2  
-\frac{\rho^{{\rm in}}_{m,\tilde m'} \rho^{{\rm in}}_{m,\tilde m} 
}{
\sum_{\tilde m''}\rho^{{\rm in}}_{m,\tilde m''}
}{T}^{(1)}_{\substack{mm'\\\tilde m\tilde m}}{T}^{(1)*}_{\substack{mm'\\\tilde m'\tilde m'}} 
\biggr)
\nonumber\\ &
+ \Od{\lambda^2}\nonumber
\end{align}
where
\[
{T}^{(1)}_{\substack{mm'\\\tilde m\tilde m'}}= \bra{m,\tilde m} \hat{T}^{(1)}\ket{m',\tilde m'}.
\]
For pure initial states, the non-analyticity in $\lambda$ of the leading contribution to the von Neumann entropy was noted in some of the early literature on momentum space entanglement in quantum field theory \cite{vanR} although the rest of the above formula is quite different from theirs. The logarithm of the coupling also appears in the computations of entanglement due to scattering of incoming pure states  \cite{seki1,Grignani:2016igg}. Here, (\ref{2ndorder}) is a generalization of that result to include all incoming states which obey (\ref{separable0}) and where $\hat{\rho}^{A{\rm in}}$ has a nontrivial kernel.

The contribution to the change of the A subsystem entropy in equation (\ref{2ndorder}) is proportional to a sum over the perturbations of the zero eigenvalues of the reduced density matrix, the leading order of which are now the second order contributions that are displayed there.  Being of second order, their signs can no longer be flipped by changing the sign of $\lambda$ or $\hat{T}^{(1)}$.  Moreover, just as we argued for the first order in perturbation theory, since they are eigenvalues of a density matrix,  each individual eigenvalue as well as the sum  over the eigenvalues must be non-negative.  
Indeed, we can explicitly confirm that the sum over the shifts of zero eigenvalues is non-negative by rewriting the summation in equation (\ref{2ndorder}) as
\begin{align}
\label{2ndorder1}
 \delta S=\,&\left[\lambda^2\ln\frac{1}{\lambda^2}\right]  \\  &\times\sum\limits_{ \substack{m\notin\kappa\\m'\in\kappa \\ \tilde m,\tilde m'}} \rho^{{\rm in}}_{m,\tilde m}   \biggl| {T}^{(1)}_{\substack{mm'\\ \tilde m \tilde m'}} 
 - \delta_{\tilde m\tilde m'}\sum_{\tilde n}
 \frac{
 \rho^{{\rm in}}_{m,\tilde n} 
 }
 {\rho^{A{\rm in}}_{m} 
}
{T}^{(1)}_{\substack{mm'\\\tilde n\tilde n}}
\biggr|^2
\nonumber\\ &
+ \Od{\lambda^2}\nonumber
\end{align}
If the right-hand-side of (\ref{2ndorder1}) is not zero, it must be positive.

What is more, the conditions for all of the terms in the summand on the right-hand-side of (\ref{2ndorder1}) to vanish are quite stringent.  It would vanish if the $\hat{T}$ matrix had zero matrix elements between all kernel and non-kernel states, $ {T}^{(1)}_{\substack{m m'\\ \tilde m \tilde m'}} =0$ if $m\in\kappa,m'\notin\kappa$. This  is certainly a logical possibility. For example, it would occur if the kernel were protected by a super-selection rule.

If it were not so, that is, if any of those $ {T}^{(1)}_{\substack{m m' \\ \tilde m \tilde m'}} \neq 0$, the summand would vanish if and only if 
\begin{align}
 {T}^{(1)}_{\substack{m m' \\ \tilde m \tilde m'}} 
&= \delta_{\tilde m\tilde m'}{T}^{A(1)}_{\substack{mm'}},&
&\text{for all }m\notin\kappa,\; m'\in\kappa
\end{align}
whence the elements of the $\hat{T}$ matrix that mix the kernel and non-kernel states act trivially, as the unit operator on the B subsystem.

 Thus, we can conclude that, if
 the criterion of equation (\ref{separable0}) is obeyed 
and the reduced density matrix $\hat{\rho}^{A{\rm in}}$ has a non-empty kernel,
the shift of entropy of subsystem A is non-negative at the order \(\lambda^2 \ln \frac{1}{\lambda^2}\). 

 If, in addition, 
 the $\hat{T}$ matrix allows transitions between non-kernel and kernel states
and, if in doing so, the $\hat{T}$ matrix acts nontrivially 
on the  B subsystem, the change in the entropy of subsystem A is guaranteed to be positive, of order $\lambda^2\ln\frac{1}{\lambda^2}$ in the weak coupling limit, and given by equation (\ref{2ndorder1}).

 \section{A simple example}
 
 The first term on the right-hand-side of equation (\ref{2ndorder1}) has a simple interpretation.  It contains the transition probability, $\left|  {T}^{(1)}_{\substack{m m'\\ \tilde m \tilde m'}} \right|^2$,  for transitions from 
initial non-zero probability states $\left|m,\tilde m\right>$ to initial zero probability states, $\left|m',\tilde m'\right>$, weighted by the probability $\rho^{\rm in}_{m,\tilde m}$ that $\left|m,\tilde m\right>$ occurs in the initial state.  

The second term on the right-hand-side of equation (\ref{2ndorder1})  compensates for some over-counting.   To appreciate that the second term is important, consider a $\hat{T}$ matrix which only acts on the $A$ subsystem, so that $$\left|  {T}^{(1)}_{\substack{m m'\\ \tilde m \tilde m'}} \right|^2= \left|  {T}^{(1)}_{\substack{m m'}} \right|^2\delta_{\tilde m\tilde m'}$$
Then the A subsystem is isolated and its time evolution is unitary.  In that case its entropy cannot change. 
We have already seen that $\delta S$ vanishes
in this case, the role of the second term in  (\ref{2ndorder}) being to cancel the entire contribution. 

As an example, consider 
a pure initial state,  $\hat{\rho}^{\rm in}=\ket{1,\tilde 1}\bra{1,\tilde 1}$. Here, $\ket{1}$ and $\ket{\tilde 1}$ could be wavepacket states of two distinct particles, for example. In this case, equation (\ref{2ndorder1}) reduces to
\begin{align}
&\delta S = \left[\lambda^2\ln\frac{1}{\lambda^2}\right] \sum_{m\neq 1,\tilde m\neq 1} 
\left|  {T}^{(1)}_{\substack{1 m\\ \tilde 1 \tilde m}} \right|^2  
+ \Od{\lambda^2}.
\label{shift of S for a pure state}
\end{align}

The result is  the total transition probability to all states with the condition that the states of both  the A and B subsystems must change. If we apply this formula to the scattering of two distinct particles, and change the normalization to entropy production rate per unit of beam flux, we find the area law of equation (\ref{area law}).  
 
 \section{When $\hat{\rho}^{A{\rm in}}$ has full rank}
 
Finally, we might ask the question about the case where the initial state has the property (\ref{separable0}) but the reduced initial density matrix $\hat{\rho}^{A{\rm in}}$  does not have a kernel, that is, $0<\rho^{A{\rm in}}_m\leq 1$ $\forall m$.  Then, the order $\lambda^2\ln\frac{1}{\lambda^2}$ term is absent from the perturbative expansion of the entropy. The dominant term is of order at least $\lambda^2$. We have
\begin{equation}\label{delta s}
\delta S = - \lambda^2 \sum_{m}\delta\rho^{A(2)}_m[\ln \rho^{A{\rm in}}_m+1]+\Od{\lambda^3} 
\end{equation}
as \(\lambda \to 0\).

Since the trace of a density matrix such as $\hat{\rho}^{A{\rm in}}$ must be unity, we must have $ \sum_{m}\delta\rho^{A(2)}_m=0$.  We can easily confirm this by examining the trace of the order $\lambda^2$ terms in the matrix $\mathcal M_{mm'}$ in equation (\ref{shift of the kernel}). We see that
\begin{align}\nonumber
&\sum_m \delta\rho^{A(2)}_m=\sum_m\mathcal M_{mm}\\
 &= \lambda^2
\sum_{\substack{m, m'  }}
\Biggl\{\sum_{\substack{ \tilde m,\tilde m' }}
[\rho^{\rm in}_{m,\tilde m}- \rho^{\rm in}_{m',\tilde m'}] \left| {T}^{(1)}_{\substack{mm'\\\tilde m\tilde m'}}\right|^2  
 \label{perteq}\\ &\hspace{3em}
-   \frac{1}{\rho^{A{\rm in}}_{m} - \rho^{A{\rm in}}_{m'}} 
\left| 
\sum_{\tilde m} [\rho^{\rm in}_{m,\tilde m}-\rho^{\rm in}_{m',\tilde m}] {T}^{(1)}_{\substack{mm'\\ \tilde m\tilde m}}
\right|^2
   \Biggr\}
   \nonumber \\\nonumber
&\hspace{2em}
+ \Od{\lambda^3}.
\end{align}
Here we used $\hat{T}^{(1)}=\hat{T}^{(1)\dagger}$. Now,
$$ \left| {T}^{(1)}_{\substack{mm'\\\tilde m\tilde m'}}\right|^2~~~{\rm and}~~~\left| 
\sum_{\tilde m}[\rho^{\rm in}_{m,\tilde m}-\rho^{\rm in}_{m',\tilde m}] {T}^{(1)}_{\substack{mm'\\ \tilde m\tilde m}}
\right|^2
$$
are symmetric under $m,\tilde m\leftrightarrow m',\tilde m'$ and  $m \leftrightarrow m' $, respectively. 
The summand in (\ref{perteq}) is therefore odd under this interchange and the sum must vanish;
$$
\sum_{m}\delta\rho^{A(2)}_m=0.
$$
Then equation (\ref{delta s}) becomes
\begin{align}\nonumber
\delta S  =& \frac{\lambda^2}{2}
\sum_{\substack{m,  m'  }}
\ln [\rho^{A{\rm in}}_{m} /\rho^{A{\rm in}}_{m'} ]
\biggl\{\sum_{\substack{ \tilde m,\tilde m' }}
[\rho^{\rm in}_{m,\tilde m}- \rho^{\rm in}_{m',\tilde m'}] \left| {T}^{(1)}_{\substack{mm'\\\tilde m\tilde m'}}\right|^2  
\nonumber \\ &
-   \frac{1}{\rho^{A{\rm in}}_{m} - \rho^{A{\rm in}}_{m'}} 
\left| 
\sum_{\tilde m}[\rho^{\rm in}_{m,\tilde m}-\rho^{\rm in}_{m',\tilde m}] {T}^{(1)}_{\substack{mm'\\\tilde m\tilde m}}
\right|^2
   \biggr\}
\nonumber \\ &+\Od{\lambda^3}\label{second last step full rank}
\end{align}
where we have used the antisymmetry of the summand to anti-symmetrize the logarithmic factor under $m\leftrightarrow m'$.  Rearranging (\ref{second last step full rank}),
\begin{align}
\nonumber
\delta S  =& \frac{\lambda^2}{2}
\sum_{\substack{m,  m' \\ \tilde m\tilde m' }}
\ln \frac{\rho^{A{\rm in}}_{m} }{\rho^{A{\rm in}}_{m'} }
[\rho^{\rm in}_{m,\tilde m} - \rho^{\rm in}_{m',\tilde m'}]   
\\ &
\;\;\times \biggl| {T}^{(1)}_{\substack{mm'\\\tilde m\tilde m'}}
-\delta_{\tilde m\tilde m'}\sum_{\tilde n}
\frac{\rho^{\rm in}_{m,\tilde n}-\rho^{\rm in}_{m',\tilde n}}{\rho^{A{\rm in}}_{m} - \rho^{A{\rm in}}_{m'}}   {T}^{(1)}_{\substack{mm'\\\tilde n\tilde n}}\biggr|^2
\nonumber \\ &+\Od{\lambda^3 }.
\end{align}

To see the indefinite sign of the change in entropy, consider a simple example of two two-level systems that is initially in the product state
$$
\hat{\rho}^{\rm in}=\biggl[x\ket{1}\bra{1}+(1-x)\ket{2}\bra{2}\biggr]\otimes \biggl[y\ket{\tilde 1}\bra{\tilde 1}+(1-y)\ket{\tilde 2}\bra{\tilde 2}\biggr]
$$
where $x,y\in[0,1]$.  Here,  
$$
\hat{\rho}^{A{\rm in}}=\biggl[x\ket{1}\bra{1}+(1-x)\ket{2}\bra{2}\biggr]
$$
has full rank when $0<x<1$. Let us 
 assume that the only nonzero components of the $\hat{T}$ matrix are
$$
 {T}^{(1)}_{\substack{12\\ \tilde 1\tilde 2}}
  ={T}^{(1)}_{\substack{12\\\tilde 2\tilde 1}}    
   ={T}^{(1)}_{\substack{21\\\tilde 1\tilde 2}}
   = {T}^{(1)}_{\substack{21\\\tilde 2\tilde 1}}=t.
$$
The result for the change in the entropy of the A subsystem is then
\begin{align}
\nonumber
\delta S  =& \lambda^2  |t|^2 
\ln \frac{x}{1-x }
[x(1-y)-(1-x)y]+\Od{\lambda^3}.
\end{align}
This result can be either positive or negative.  It changes sign as $y$ varies from zero to one, irrespective of the value of $x$, as long as $x\neq 0,\frac{1}{2},1$. 
 
 Another example is the situation where subsystem A is in a thermal state and subsystem B is in a pure state,
 $$
 \rho^{\rm in}_{m\tilde m}= \frac{e^{-\beta E_m}}{\sum\limits_{E} e^{-\beta E}}\delta_{\tilde m,\tilde 1}
 $$
 The thermodynamic entropy of the A system is equal to its von Neumann entropy. 
 
 Using equation (\ref{second last step full rank}) to compute the change of entropy due to scattering yields
 \begin{align}
\nonumber
\delta S  =&  \lambda^2 
\sum_{\substack{m,  m' \\ \tilde m\tilde m' }}
\beta(E_{m'}-E_{m})\frac{e^{-\beta E_m}}{\sum\limits_{E} e^{-\beta E}} 
\biggl| {T}^{(1)}_{\substack{mm'\\ \tilde 1\tilde m'}}\biggr|^2
 +\mathcal O(\lambda^3)
\end{align}
Now we can see that the overall sign of the entropy shift will depend on the relative weights of $E_{m'}$ and $E_{m}$ in the summation and the states of
the B system can be set so that it is either positive or negative. 
 
For example, we could assume that the $\hat{T}$ matrix conserves energy for the process $m+\tilde 1\to m'+\tilde m'$, that is, that
$E_m+E_{\tilde 1}=E_{m'}+E_{\tilde m'}$.  In that case, we 
can replace the above with 
 \begin{align}
\delta S  =& - \lambda^2 
\sum_m \frac{e^{-\beta E_m}}{\sum\limits_{E} e^{-\beta E}} 
\sum_{\substack{ m' \\ \tilde m\tilde m' }}
\beta(E_{\tilde m'}-E_{\tilde 1})
\biggl| {T}^{(1)}_{\substack{mm'\\ \tilde 1\tilde m'}}\biggr|^2
 +\mathcal O(\lambda^3)
 \label{thermal}
\end{align}
The sign of the right-hand-side now depends on the relative magnitudes of the term with $E_{\tilde m'}$ and the term with $E_{\tilde 1}$. 

If, for example, $\tilde 1$ is the lowest energy state of the B subsystem, so that $E_{\tilde 1}$ is the lowest possible energy of the B subsystem, the right-hand-side of (\ref{thermal}) is negative and entropy of the A system decreases.  This is the onset of thermalization, where a thermal state of the A subsystem at temperature $1/k_B\beta$ is in weak contact with a zero temperature state of the B subsystem. 
The thermal state begins to cool, decreasing its thermal entropy. 

In order for the entropy of the A subsystem to increase, $E_{\tilde 1}$ in (\ref{thermal}) would have to be a high energy state so that typical terms which contribute to the sum have $(E_{\tilde m'}-E_{\tilde 1})<0$.  Then the weak contact of the two subsystems would be the onset of thermalization where the temperature and therefore the entropy of the A subsystem would begin to increase. 

We see that, when the incoming density matrix of the A subsystem has full rank, even simple product incoming states of the composite system can lead to A subsystem entropy increasing or decreasing, depending on the  states of the B subsystem.  We suspect, but we have not proven that this is always the case and that, when the incoming A subsystem density matrix has full rank, the $\hat T$ matrix and B subsystem states can always be organized so that the A subsystem entropy increases or decreases.  An exception may be the state of maximum entropy where $\rho^{A{\rm in}}$ is proportional to the unit matrix.  In the infinite dimensional Hilbert spaces that are typical of scattering theory such a state would have to be defined by a subtle limiting process.

 \section{Conclusion}
 
 In conclusion, we restate our central result, that in perturbation theory, the entropy of a subsystem of a bipartite quantum system will be non-decreasing at order \(\lambda^2 \ln \frac{1}{\lambda^2}\) if the initial reduced density matrix of that subsystem has a nonempty kernel and if the initial density matrix of the composite system commutes with the projectors onto the eigenspaces of the
 reduced density matrix of the A system, that is, if equation (\ref{separable0}) or equivalently (\ref{special0}) is satisfied. 
 
 In addition, the entropy will be strictly increasing if the above conditions hold and if,  $\hat{T}^{(1)}$ allows for transitions in and out of the kernel of \(\hat{\rho}^{A{\rm in}}\) and if $\hat{T}^{(1)}$ acts nontrivially on the B subsystem.

  \noindent
  Some of the technical results in this paper were the content of the thesis \cite{dm} of one of the authors (DM). 
This work is supported in part by NSERC. One of the authors (GWS) thanks the Galileo Galilei Institute for Theoretical Physics for the hospitality and the INFN for partial support during the completion of this work.

\end{document}